\documentclass{jfm}
\usepackage{lineno}

\usepackage{color}
\definecolor{myblue}{rgb}{0,0,1}
\definecolor{myyellow}{rgb}{1,1,0}
\definecolor{myred}{rgb}{1,0,0}
\definecolor{mygreen}{rgb}{0,1,0}
\usepackage{soul}
\usepackage{tikz}
\usetikzlibrary{shapes}

\newcommand{\markertri}{\raisebox{0pt}{\tikz{\node[draw,scale=0.45,regular
      polygon  sides=3,rotate=0](){};}}}

\newcommand{\markercirc}{\raisebox{0.6pt}{\tikz{\node[draw,scale=0.5,circle](){};}}}
\usepackage{pgf}
\usepackage{graphicx}
\usepackage{epstopdf, epsfig}
\usepackage{amsmath}
\usepackage{url}

\newcommand{\add}[1]{\textcolor{black}{#1}}
\newcommand{\remove}[1]{{#1}}
\newcommand{\modify}[1]{\textcolor{black}{#1}}

\shorttitle{On the detection of interfacial layers}
\shortauthor{D.S. Fan,  J.L. Xu, M.X. Yao and J.-P. Hickey}

\title{On the detection of internal interfacial layers in turbulent flows}

\author{Duosi Fan\aff{1,2},
  Jinglei Xu\aff{1,2}, \textcolor{black}{Matthew X. Yao}\aff{2} 
 \and Jean-Pierre Hickey\aff{2}   \corresp{\email{j6hickey@uwaterloo.ca}}}

\affiliation{
\aff{1}School of Energy and Power Engineering, Beihang University,
Xueyuan Road, Beijing 100191, China
\aff{2}Department of Mechanical and Mechatronics Engineering, University of
Waterloo, University Avenue, Waterloo N2L 3G1, Canada
}

\begin{document}

\maketitle

\begin{abstract}
A novel approach to identify internal interfacial layers, or IILs, in wall-bounded turbulent flows is proposed. Using a Fuzzy Cluster Method (FCM) on the streamwise velocity component, a unique and unambiguous grouping of the Uniform Momentum Zones is achieved, thus allowing the identification of the IILs. The approach overcomes some of the key limitations of the histogram-based IIL identification methods. The method is insensitive to the streamwise domain length, can be used on inhomogeneous grids, uses all the available flow field data, is trivially extended to three dimensions, and does not need user-defined parameters (e.g. number of bins) other than the number of zones\add{. The number of zones can be automatically determined by an \emph{a priori} algorithm based on a Kernel Density Estimation algorithm, or KDE.} The clustering approach is applied to the turbulent boundary layer (experimental, planar PIV) and channel flow (numerical, DNS) \add{at varying Reynolds numbers}. \modify{The interfacial layers are characterized by a strong concentration of spanwise vorticity, with the outer-most layer located at the upper edge of the log-layer. The three-dimensional interface identification reveals a streak-like organization; we show that the organization of the IILs is  correlated to the underlying wall-bounded turbulent structures. }

\end{abstract}


\section{Introduction}
The delineation between regions of different turbulence intensity is central to the study of turbulent flows. The most obvious boundary occurs at the interface between turbulent and non-turbulent regions in many free shear or wall-bounded flows such as jets, wakes, mixing, and boundary layers. As the turbulent/non-turbulent interface (TNTI) bounds the turbulent region, the exchange of mass, momentum, and energy between the turbulent and non-turbulent regions must occur through this interface. Consequently, the evolution of the flow properties can be understood and explained through the prism of the dynamics of the interfacial layer, as done in the seminal work by \citet{corrsin1955free} and many others since (see e.g. \cite{bisset2002turbulent,westerweel2005mechanics,Silva2014}).

Although not as obvious, an interfacial layer is also observed in wall-bounded turbulent flows. \citet{meinhart1995existence} reported the existence of an internal interfacial layer (denoted here as IIL) inside the turbulent boundary layer which delineates the zones with relatively uniform streamwise momentum, also called Uniform Momentum Zone or UMZ. IILs play an important role in momentum transport as the spanwise vorticity is localized near the interface which results in a rapid change in the instantaneous streamwise velocity. Jumps in streamwise velocity are a common feature to both the TNTI \citep{bisset2002turbulent,chauhan2014turbulent} and IIL \citep{de2017interfaces,kwon2014quiescent}. The existence of UMZs and IILs is also consistent with the prevailing understanding of coherent structures within wall-bounded turbulent flows. \citet{adrian2000vortex} indicated that coherent alignment of hairpin vortices in the streamwise direction (or vortex packets) creates the appearance of UMZs. Recently, a representative model for large scale motion (LSM) presented by \citet{saxton2017coherent}
also results in a zonal-like structure of the turbulent boundary layer. %

Although interfacial layers are central to our fundamental understanding of turbulent flow dynamics, the identification approaches--particularly for the IIL--are often ambiguous, arbitrary and plagued with interpretational uncertainties. A user-defined threshold on the magnitude of the enstrophy is widely accepted as a means to identify the TNTI in turbulent wake \citep{bisset2002turbulent}, boundary layer \citep{borrell2016properties} and other flows \citep{holzner2011laminar} as the turbulent and non-turbulent regions are rotational and irrotational, respectively. On the other hand, as IILs are embedded within turbulent--thus rotational--flow, the vorticity-based thresholding methods are ineffective at detecting the bounds of the  UMZs. \citet{eisma2015interfaces} identified IILs in a turbulent boundary layer and attempted to eliminate the influence of mean shear by a triple decomposition of vorticity following  \citet{kolar2007vortex}. However, the threshold in their investigation is nonetheless selected empirically. \citet{adrian2000vortex} suggested that the local peak in streamwise velocity histogram identifies the modal velocity representing a uniform momentum zone; the iso-surface of the streamwise velocity at the local minimum between the peaks corresponds to the interfacial layer. Following \citet{adrian2000vortex}, \citet{de2016uniform} concluded that the number of UMZs  increases log-linearly with the Reynolds number and the interface is characterized by an instantaneous streamwise velocity jump. However, the histogram peaks are sensitive to the number of bins and streamwise domain ($L_x$) and selected flow region under consideration (many approaches neglect the near-wall and freestream data to construct the histogram). An incorrect selection of the number of bins or streamwise domain may lead to incorrect peaks in the histogram. \citet{adrian2000vortex} suggested that $L_x \approx \delta$, where $\delta$ is the thickness of boundary layer, is needed for peak identification in the histogram method but \citet{de2016uniform} argued that this quantity should be selected based on inner scaling parameters. They determined $L_x^+\approx 2000$ was appropriate for turbulent boundary layers and \(L_x^+\) introduces  an implicit filter on the histogram. UMZs have also been reported in turbulent channel flows \citep{kwon2014quiescent} where $L_x=1.2h$  (with $h$ being the channel half-height) was selected after examining the modal velocity change with increasing $L_x$. Furthermore, the histogram approach \remove{cannot} \add{cannot trivially}  be extended to  three-dimensional interface identification as the spanwise corrugations of the velocity histogram act to smear out the visible peaks in the two-dimensional planar field of view.

We propose a unified approach to identify interfacial layers in two- or three-dimensional turbulent flow fields--obtained experimentally or numerically--using a Fuzzy Cluster Method (FCM). This approach is repeatable, robust, unambiguous and can be used for either TNTI or IIL with varying levels of spatial resolution and/or grid spacing inhomogeneities. The novel clustering approach identifies the  same IIL as the histogram-based identification method, but overcomes many of the limitations of the latter method. Most importantly it can provide a full three-dimensional description of the internal interface which allows for a more detailed  investigation of the conditional flow statistics at the boundaries of the  UMZs.  This work analyzes publicly available planar PIV and three-dimensional direct numerical simulation (DNS) data and our code is provided online for full repeatability and transparency of the obtained results. In this work, the details of the procedure are presented in section \ref{sec:detection} and  the results is summarized in section 3. The FCM is applied to experimental turbulent boundary layer  data (section \ref{sec:valid}) and to numerical three dimensional turbulent channel flow  data (section \ref{sec:3d}). The increased statistical sample size of the three-dimensional data provides the opportunity to study the IILs with detailed conditional statistics and topological feature extraction. \add{The sensitivity and robustness of the proposed method is investigated in section \ref{sec:robustness}; it is shown that the method is robust against perturbations to the initial conditions, the initialization of the membership functions, and the \emph{a priori} selection of the number of zones.}

\section{Detection of interfacial layers based on cluster analysis}\label{sec:detection}

\add{The following sub-section describes the principles of a Fuzzy Clustering Method (FCM) called \emph{fuzzy c-means clustering} which is applied to the identification of internal zones. An additional method is also proposed, in subsection 2.2, to automatically determine the number of zones.}
\add{\subsection{Fuzzy Cluster Method}}
A novel approach is proposed which identifies the internal interfacial layers by grouping contiguous and non-contiguous zones according to common flow features. Here, we consider flow fields from an instantaneous snapshot of a large-eddy (LES), direct numerical simulation (DNS) or  particle image velocimetry (PIV,  either planar or volumetric) database. The entirety of the flow data within a given snapshot forms a sequence of observations, \(\mathbf{\Phi} = \{\Phi^m\}_{m=1}^{N}\), where \(N\) represents the total number of grid points in the snapshot and $\Phi$ a scalar quantity such as pressure or velocity component  \citep{friedman2001elements}. As the clustering is done on the streamwise velocity in the current paper, we replace \(\Phi\) by \(U\) to simplify the notation. The cluster analysis aims at grouping the observations \(\{U^m\}_{m=1}^N\) into \(K\) clusters, \(\Pi_k\) (with $k=1,\cdots,K$) based on their similar streamwise velocity. To quantify similar flow features, a distance metric between two observations is defined as \(D_{mn}=|U^m-U^n|\), where the norm is in the Euclidean space. We adopt a Fuzzy Cluster Method (FCM) \citep{dunn1973fuzzy,bezdek1981objective} in which a membership coefficient \(u_{mk}\) represents the probability of the observation \(U_m\) belonging to cluster \(\Pi_k\). Based on this definition:  \(\sum_{k=1}^{K}u_{mk}=1\).  We define \(c^k\) as the centroid of cluster \(k\), accordingly, it represents the characteristic velocity of the cluster. It corresponds to the average of all observations with the membership function \(u_{mk}\):
\begin{linenomath*}
  \begin{equation}
    \label{eq:centroid}
    c^k = \frac{\sum_{m=1}^{N}(u_{mk})^pu_m}{\sum_{m=1}^{N}(u_{mk})^p}
  \end{equation}
\end{linenomath*}
where \(p\in [1,\infty)\) is the fuzziness parameter in FCM and is usually set to \(2\) in practice \citep{Pal:1995:CVF:2234581.2235722}. The cluster centroids form a centroid vector \(\vec{\mathbf{c}}=(c^1,\dots,c^K)\). The cluster centroid represents a characteristic velocity of each cluster and as for the modal velocity in the histogram method, represents the characteristic velocity of each uniform momentum zone. Once \(\vec{\mathbf{c}}\) and \(u_{mk}\) are determined, the variance can be defined as an object function
\begin{linenomath*}
  \begin{equation}
    \label{eq:object}
    J_p(\mathbf{U},\vec{\mathbf{c}}) = \sum_{m=1}^{N}\sum_{k=1}^{K}(u_{mk})^p(D_{mk})^2.
  \end{equation}
\end{linenomath*}
The optimal \(\vec{\mathbf{c}}\) is determined by the minimization of the objective function via standard Lagrange multiplier method. As a result, the FCM is a bootstrapping algorithm which is summarized as follows: 
\begin{itemize}
\item[(1)]Initialize the membership coefficient \(u_{mk}\) for each observation; 
\item[(2)]~Compute the centroid of the clusters \(c^k\) with equation \eqref{eq:centroid}; 
\item[(3)]~Recompute the membership coefficient using the new \(c^k\)
\begin{equation}
u_{mk}=\left({\sum_{j=1}^{K}\left({D_{mk}}/{D_{mj}}\right)^{{2}/{(p-1)}}} \right)^{-1};
\end{equation}
 \item[(4)]~Execute steps 2 and 3 iteratively until \(|\vec{\mathbf{c}}-\vec{\mathbf{c_0}} | \leq \epsilon \) where \(\epsilon\)  is a pre-determined error threshold and \(\vec{\mathbf{c_0}}\) is the centroid vector in the previous iteration. 
\end{itemize}

Once converged, we assign the observation \(U_m\) to \(\Pi_k\) with the highest membership \(I_m=\operatorname{arg\,max}_{\tiny {k}}(u_{mk})\). Although it is reported that FCM can be sensitive to initial \(u_{mk}\) \citep{pham2000current}, the present results are shown to be robust and invariant to the initialization. Naturally, the FCM algorithm requires a user-defined number of clusters \(K\) and various approaches to determine the optimal \(K\) are possible. For example, \citet{bezdek1974numerical} defined a partition coefficient to measure the ``overlap" between clusters and \(K\) is selected when the partition coefficient reaches a minimum. \add{In the following section, we propose an algorithm, based on a Kernel Density Estimation (KDE) approach,  to automatically select the appropriate number of zones for a given dataset.} The interface identification code \add{and the method to identify the number of zones}--both written in Python--as well as all the datasets used are freely accessible online. The code is accessible at \url{zenodo.org} with the DOI: (to be set after publication).

\add{\subsection{Selection of the number of IILs}}
\add{The clustering approach rests on the understanding that we have an \emph{a priori} knowledge of the number of UMZ or IIL in a given flow. At low-Reynolds number, the number of UMZs can be inferred from a physical understanding of the various regions within wall-bounded turbulent flow. This simplistic approach breaks down when faced with higher-Reynolds number wall-bounded turbulent flows for which the number of zones are dependent on the Reynolds number \citep{de2016uniform}. For this reason, we propose an alternate clustering approach that automatically determines the optimal number of zones based on the same streamwise velocity parameter as the FCM. The approach, which rests on a Kernel Density Estimation (KDE) algorithm, does suffer from some of the same shortcomings as the histogram-based approach (requires an adequate selection of domain length, inability to account for near-wall grid stretching etc.); in the current implementation it is only used to determine the number of UMZs, and by extension, the number of IILs.  }

\add{Kernel Density Estimation is a statistical method which is used to estimate the underlying probability density function from a set of sampled data. A kernel is placed at each observation and the results are summed to produce a smooth probability density function which accounts for the density of the observed samples. Two important parameters of the KDE are the shape of the kernel and the bandwidth, which acts as a smoothing parameter. Although there are a variety of possible kernels which can be used, the bandwidth has a much more significant effect on the results of the KDE. In our analysis, we use a Gaussian kernel; more information on the density estimation can be found in \citet{silverman1998density}. Based on a minimization of the mean integrated squared error, the optimal bandwidth $h$ for a normal kernel is calculated as \citep{scott1992multivariate}:}
\begin{equation}
  h = \sigma \left(4/3 \right)^{1/5} n^{-1/5}
\end{equation} 
\add{where $\sigma$ is the standard deviation and $n$ is the number of observations. }

\begin{figure}
  \centerline{\includegraphics[width=0.7\linewidth]{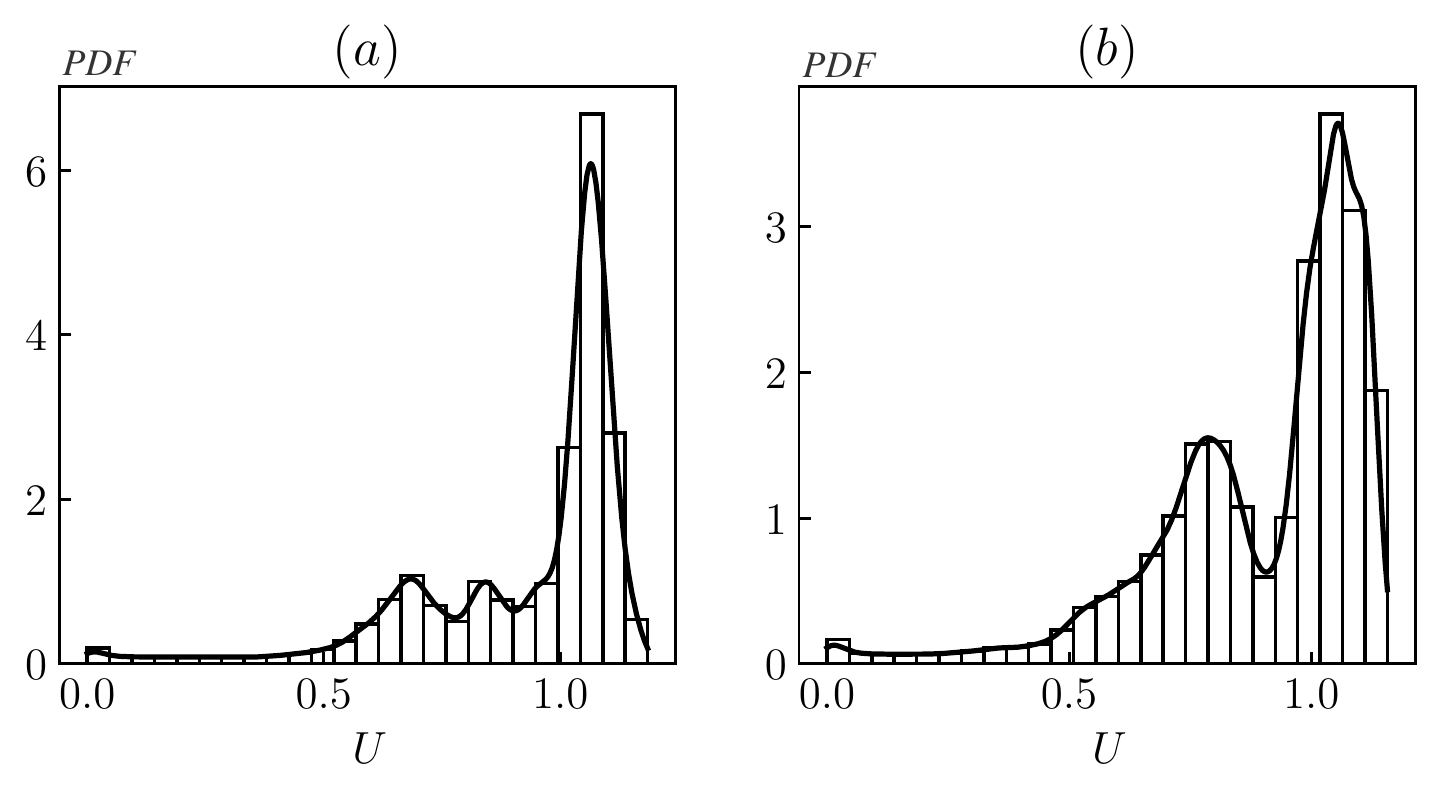}}
  \caption{Kernel Density Estimation at two different streamwise locations of DNS channel flow data \citep{lee_moser_2015}. }
  \label{fig:KDE}
\end{figure}

\add{An example of a KDE applied over the same window size at two different  streamwise locations of DNS channel flow data from \citet{lee_moser_2015} is shown in figure~\ref{fig:KDE}. The automatic bandwidth determination allows for an accurate identification of the number of internal zones.  Similar to the histogram method, the KDE is conducted on a slice-by-slice basis. Although the number of peaks is defined without user input, the clustering would subsequently have to be performed on the thin region where the predicted number of zones is valid. }

\add{To group the data, a clustering approach based on KDE called mean shift clustering can be applied. Similar to KDE, the correct bandwidth selection has been studied and there are many recommendations, eliminating the need for the user to arbitrarily input a bandwidth. Despite promising results on PIV data, which has well defined peaks, the mean shift clustering approach misidentifies the peaks in the presence of the smaller grid sizes near the wall in DNS results. The newly proposed Fuzzy Clustering Method, presented in the previous subsection, does not have the same limitations.}

\add{Although the selection of the number of zones is a key input to the fuzzy clustering method, the correct identification of IIL is nonetheless possible if the selected number of zones is slightly incorrect (say +/- 1), see section \ref{sec:robustness}, which discusses the sensitivity and robustness of our method to the selection of the number of zones. In other words, if a flow is composed of three UMZs (two IILs), the same two IILs will be identified by the method if we wrongly input three IIL. This robustness assures a correct identification of the dynamically significant parts of the flow. }\\

\section{Result and discussion} \label{sec:result}
\subsection{Experimental and numerical databases}\label{sec:experiment}
\add{Three} publicly accessible turbulent flow databases -- one experimental (PIV), \add{two} numerical (DNS) --  are used for the validation of the proposed method. Combined with our online accessible cluster algorithm, this assures a repeatable, transparent, and generalizable  assessment of our approach.  First, we compare our cluster identification method against \citet{adrian2000vortex}'s histogram-based approach by examining the planar PIV dataset of a turbulent boundary layer by \citet{tomkins1998structure}.  Second, the organization and conditional sampling of the IILs are analyzed using two snapshots of the incompressible channel flow DNS by \citet{del2004scaling} \add{and one  snapshot by \citet{lee_moser_2015}}. The friction velocity Reynolds number of the boundary layer  (where \(h\) is boundary layer thickness) is \(Re_{\tau}=u_{\tau}h/\nu=2216\). The snapshots of the channel flow (\(h\) is channel half height) are at 550, 950 \citep{del2004scaling},  \add{and 5186} \citep{lee_moser_2015}. The extent of the spatial domain and grid resolution of the datasets are summarized in table \ref{tab:piv};  further details can be found in \add{the corresponding references}.

\begin{table}
 \begin{center}
\def~{\hphantom{0}}
 \begin{tabular}{ccccccccl}
  & \(\Rey_{\tau}\) & \(L_x\) &\(L_y\) &\(L_z\) &\(\Delta x^+\) &\(\Delta y^+\)&\(\Delta z^+\) & References\\ \\
   Turb. BL PIV & 2216&  \(2.9\delta\)& \(1.2\delta\) & -& 30& 30 & - &\citet{tomkins1998structure}\\
   \\
  Channel DNS & 550& \(8\pi h\)& \(2h\)& \(3\pi h\)&
                                                     8.9&\(0.018-4.44\)&
                                                                         3.3 & \citet{del2004scaling}
   \\
   & 950& \(8\pi h\)& \(2h\)& \(3\pi h\)&
                                                     7.6&\(0.031-7.62\)&
                                                                         3.8 & \\ \\
   \add{ Channel DNS} & \add{5186} & \(8\pi h\)& \(2h\)& \(3\pi h\)& 12.7&\(0.498-10.3\)&6.4 & \add{\citet{lee_moser_2015} }
                                                                                                                                   
 \end{tabular}
 \caption{Spatial domain and grid resolution of the boundary layer PIV
   measurements and the channel flow simulations.}
 \label{tab:piv}
\end{center}
\end{table}

  
\begin{figure}
  \centerline{\includegraphics{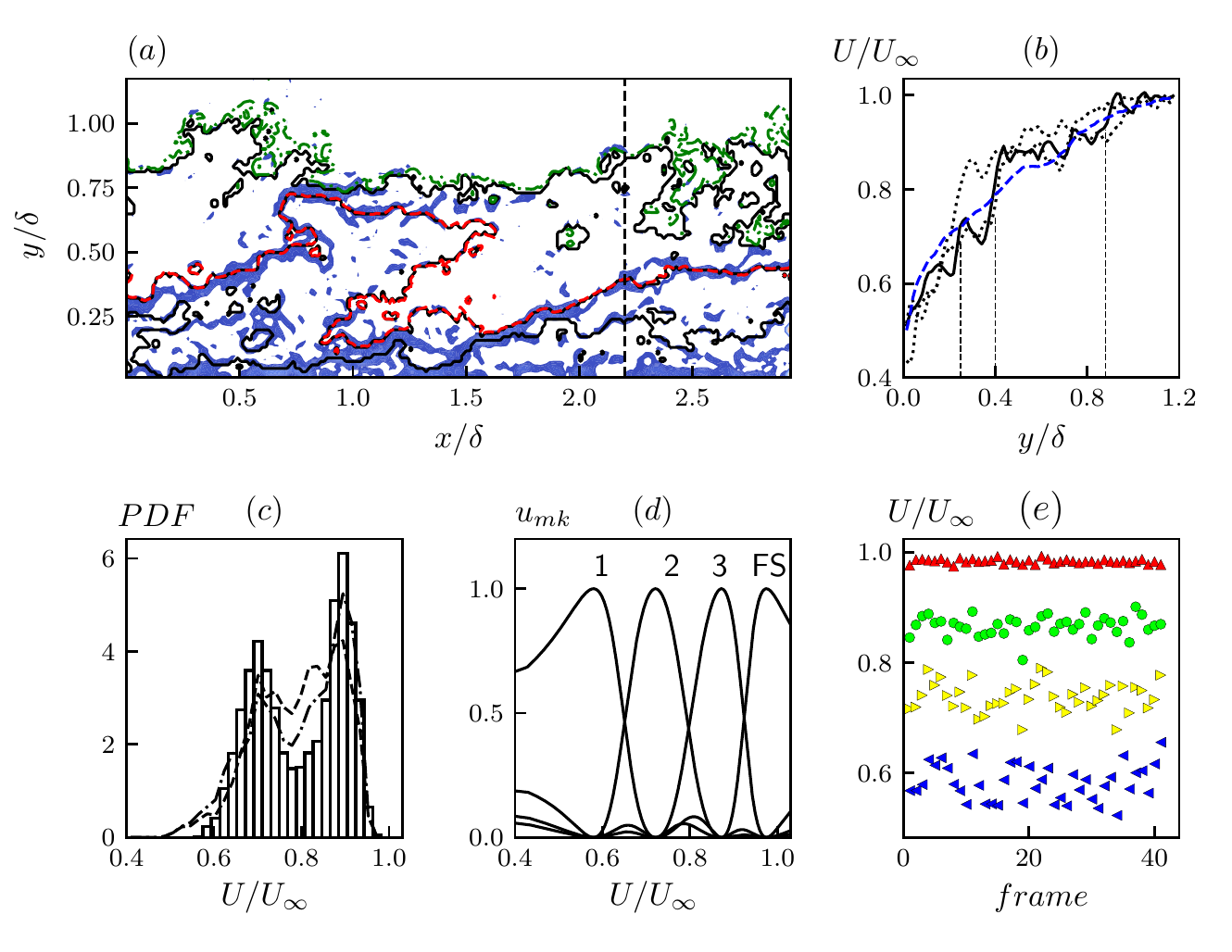}}
  \caption{(a) Comparison of the following interfacial layer identification algorithms on a turbulent boundary layer PIV frame:  cluster-based algorithm  (solid, black lines),  histogram-based IIL (dashed, red line) and threshold-based TNTI (solid green line).   The approach by \citet{de2016uniform} is used for the IIL detection. For the TNTI,  a threshold on the local turbulent kinetic energy is used as proposed by \citet{chauhan2014turbulent}. The magnitude of the local spanwise vorticity is overlaid \add{(minimum threshold on the non-dimensionalized vorticity magnitude is set to $10^{-4}U_{\infty}/\delta$ to highlight the vorticity concentration)}. (b) Instantaneous streamwise velocity profile (solid line) at \(x/\delta=2.2\) (shown as a dashed vertical line in (a)) and mean velocity computed over the PIV frame (dashed, blue line). Two randomly selected instantaneous streamwise velocity profiles are overlaid (dots). (c) Histogram of streamwise velocity based on streamwise domain  length of  \(L_x^+=2000\) (bins), \(4000\) (dashed-line) and \(6000\) (dashed  dotted-line). (d) Cluster membership function, \(u_{mk}\), for  zones 1, 2, 3 and FS.  (e) Cluster centroid in all available PIV frames (41 in total) using FCM}
 
  \label{fig:illustration}
\end{figure}

\subsection{Internal interface in the turbulent boundary layer}\label{sec:valid}
We first compare the IIL identification using the histogram and clustering approach on the planar PIV data of a turbulent boundary layer. The  algorithm proposed by \citet{de2016uniform} was used to identify the modal velocity of the UMZs based on the histogram peaks of the streamwise velocity. Two key features of \citet{de2016uniform}'s approach are: (1) the vector field above the TNTI and below \(y^+=100\) is excluded from the histogram; (2) the histogram is constructed on a constrained domain of \(L_x^+=2000\).  These imposed constraints on the histogram calculation are well reported in the literature. The  removal of the near wall and irrotational flow regions is needed to clearly observe multiple peaks in the histogram distribution, otherwise skewed modal velocities and/or additional peaks are observed.  Furthermore, previous PIV studies on IILs \citep{de2017interfaces} only cover a limited field of view (typically \(2h\)), primarily due to the ambiguity of the histogram-based method over a large spatial extent despite that IILs have been shown to correlate with very large scale motion (VLSM) having a streamwise length proportional to \(h\) \citep{adrian2000vortex}. \add{Although the footprint of these VLSM are expected to be seen in the IIL, the limited field of view renders the detailed study of the interaction of these structures with the interfacial layers more difficult.}

The proposed clustering algorithm overcomes the above-noted limitations of the histogram-based approach.    \modify{The FCM only requires the \emph{a priori} knowledge of the number of UMZs. The KDE algorithm,  described in section~\ref{sec:detection}.2,  correctly identifies four UMZs without the need of any user input. The identified zones have a direct physical representation in this flow.} Near the wall, in the viscous sub-layer, the viscosity effects dominate the dynamics and the length scales are proportional to \(\nu/u_{\tau}\) (zone 1).  In zone~2,  the flow is still influenced by the near-wall turbulence and inner wall scalings but the relative importance of the viscosity is reduced; this usually corresponds to the logarithmic region in the mean velocity profile.  At the outer edge of the boundary layer (zone~3), the velocity profile is defined by outer scaling parameters, typically \(\delta\). Finally, the free-stream region of the boundary layer is non-turbulent and irrotational (zone~4 or FS). We note that the TNTI is identified at the interface between zones~3 and 4. Although the threshold-based TNTI identification methods  are simple and straightforward, the cluster algorithm implicitly delineates this interface. The present cluster-method can be used for TNTI identification when, for example, coarse experimental data at the outer boundary layer edge limit the ability to compute the local vorticity needed for the thresholding identification.

The direct comparison of histogram-based and cluster interface identification approaches is presented in figure \ref{fig:illustration}(a), which corresponds to the PIV frame in \citet{adrian2000vortex} (figure 19 in their paper). The streamwise velocity histogram has a local minimum at \(U/U_{\infty}=0.8\) (when \(L_x^+=2000\)), see figure \ref{fig:illustration}(c), which corresponds to the IIL delineating uniform momentum zones. In the cluster-based approach, we include all velocity vectors in the PIV frame and obtain the same interfaces (figure \ref{fig:illustration}(a)) and modal velocities (figure \ref{fig:illustration}(c) and (d)). The inspection of the instantaneous streamwise velocity profile at one $x/\delta$ location in  figure \ref{fig:illustration}(b), reveals a step-like change at the identified interfacial layers; similar step-like shifts are observed at the interfaces at other streamwise locations (not shown).  As the IILs are conceptually thought to represent thin shear layers \citep{meinhart1995existence}, we note a vorticity concentration in their vicinity  (figure \ref{fig:illustration}(a)); the vorticity  is observed at all of the three interfacial layers separating the four zones of the flow.

As noted by \citet{de2016uniform}, the histogram-based IIL identification method is sensitive to the streamwise domain under consideration. By increasing the domain length from \(L_x^+=2000\) to  4000, a  third peak emerges in the histogram,  see figure \ref{fig:illustration}(c); this additional peak vanishes when the domain length is extended to \(L_x^+=6000\). Additionally, by considering the data points near the wall  (\(y^+<100\)) or in the free stream, the streamwise velocity histogram cannot uniquely identify the IIL on this dataset. Despite these constraints and limitations of the histogram method, the unambiguous identification of histogram peaks is only successful in approximately half of the PIV frames in the  database  (41 planar PIV frames are available). In many frames,  clear modal peaks cannot be deciphered. However, hairpin structures and vortex packets  can be observed in most frames. If we accept that UMZs are the result of these structures, UMZs and IILs should exist even when the histogram method fails. The modal velocities computed from the cluster-based method shows an expected  frame-to-frame  variation but a high consistency,  especially for zones 3 and FS, is noted among all PIV frames as shown figure \ref{fig:illustration} (e). The lower resolution near the wall may  account for the relatively higher variation in modal velocities of zones 1 and 2.

\subsection{Three-dimensional properties of interfacial layer in a turbulent channel flow}\label{sec:3d}
\begin{figure}
  \centerline{\includegraphics{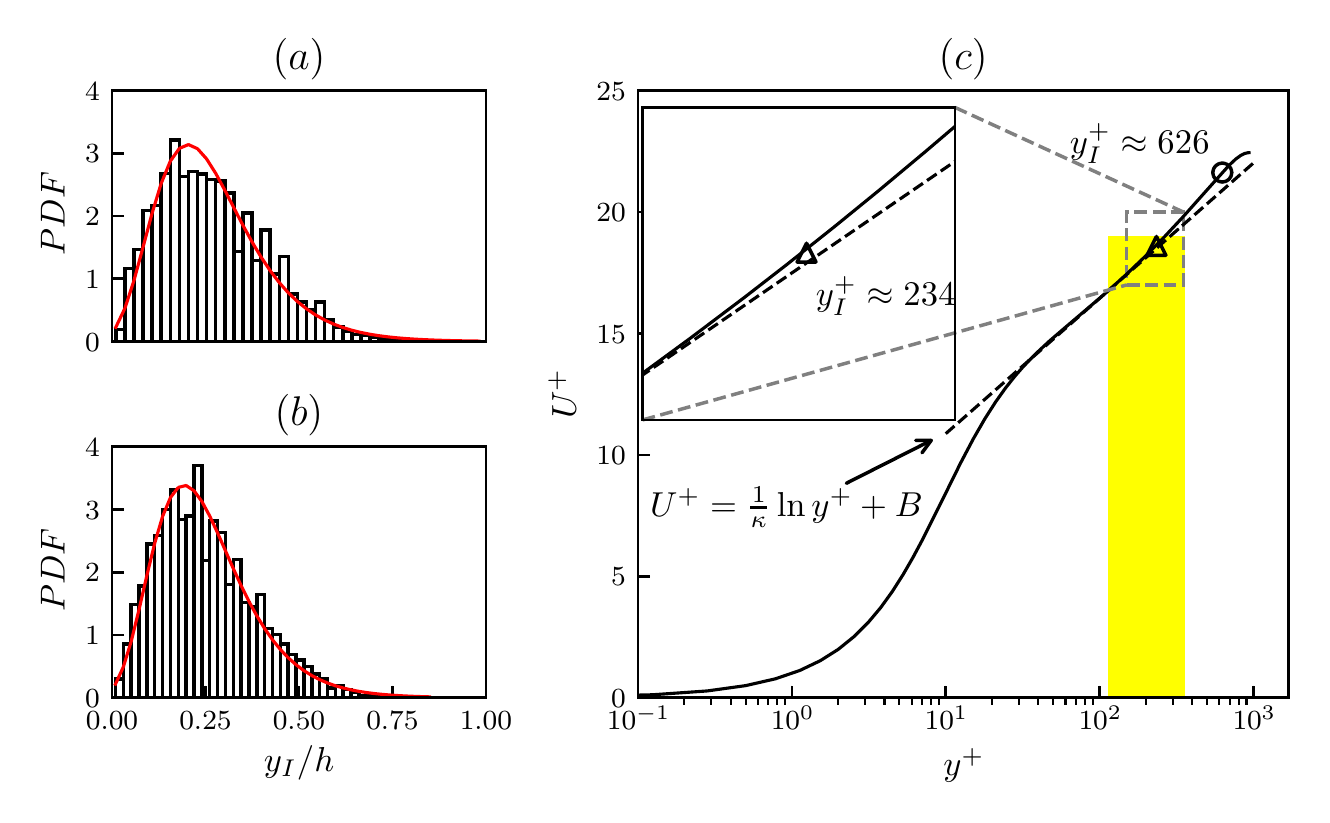}}
  \caption{The distribution of the height of the IIL (between zones 2 and 3) for a turbulent channel flow at \(Re_{\tau}=550\) (a) and \(950\) (b). The solid, red lines correspond to a \add{fitted log-normal probability distribution function}. (c)  Mean streamwise velocity profile (normalized in wall units) is shown relative to the mean height of IIL between zone 2 and 3 for \(Re_{\tau}=950\) case. The log law \(U^+=(1/\kappa)\ln(y^+)+B\) (dashed line) is obtained by a linear  fit of data between \(y^+=3Re_{\tau}^{1/2}\) and \(0.15Re_{\tau}\) (\(\kappa =0.41\)) following \citet{marusic_monty_hultmark_smits_2013}. The mean height is highlighted by  (\protect\markertri) and (\protect\markercirc) corresponds to height of the quiescent core interface of \citet{kwon2014quiescent}. \add{The width of the vertical bar denotes the standard deviation of the  interface height.}}
  \label{fig:distribution}
\end{figure}

As shown in the previous subsection, the TNTI and IILs are unambiguously identified with a robust and repeatable cluster algorithm in a turbulent boundary layer. The interfaces obtained by clustering match those from the histogram  (IIL) and turbulence threshold (TNTI) methods while overcoming many of the limitations of these classical approaches.  Despite their three-dimensional nature, the internal interfacial layers have historically been identified from two-dimensional, experimental data. Unlike the TNTI identification which is trivially extended to three-dimensions  \citep{bisset2002turbulent,borrell2016properties, hickey_hussain_wu_2013}, a surface IIL identification on volumetric data suffers from conceptual and practical limitations when using a histogram-based approach.  As highlighted by \citet{kwon2014quiescent} (figure 3(f)), the modal velocity peaks (used to identify the UMZs) are not detected in the streamwise velocity histogram in the three-dimensional channel flow.  Only  the very prominent modal velocity of the quiescent core is observed.  Although the velocity histogram helps in the identification of the quiescent core, the detection of interfacial layers is impeded. Additionally, if we consider all numerical grid points, the local grid refinement near the wall gives rise an artificial peak in the histogram due to a larger cell count of near-wall, lower velocity points. The current cluster-based approach can be extended to unambiguously identify these three-dimensional, internal interfacial surfaces from numerical databases.  In this section, we apply the cluster algorithm to three-dimensional,  incompressible channel flow DNS data. Based the extracted interfaces, we study the geometric properties and the conditional average flow field about this internal \add{interface}.

Unlike the turbulent boundary layer which is typically bound by an irrotational, free-stream region, the turbulent channel flow is fully turbulent. Based on this, we assume that the channel flow consists only of three UMZs: a viscous-dominated near-wall region (zone 1), a wall turbulence-dominated region (zone 2) and an outer scale-dominated region (zone 3). Zone 3 can also be called the \emph{``quiescent core"}  although the exact definition used here differs somewhat from  \citet{kwon2014quiescent}. \add{This zonal delineation is supported by the results of the automatic KDE-based algorithm to determine the number of zones which is presented in section~\ref{sec:detection}. } The interfacial layer between zones 2 and 3 is of particular interest as it delineates the largest UMZs.  As the structure of this interface is highly convoluted, we simplify the IIL geometry to only consider the outer hull, thus overlooking any overhanging regions. This simplification is used for the height distribution, correlations, and conditional average computations in this subsection.

\begin{figure}
  \centerline{\includegraphics{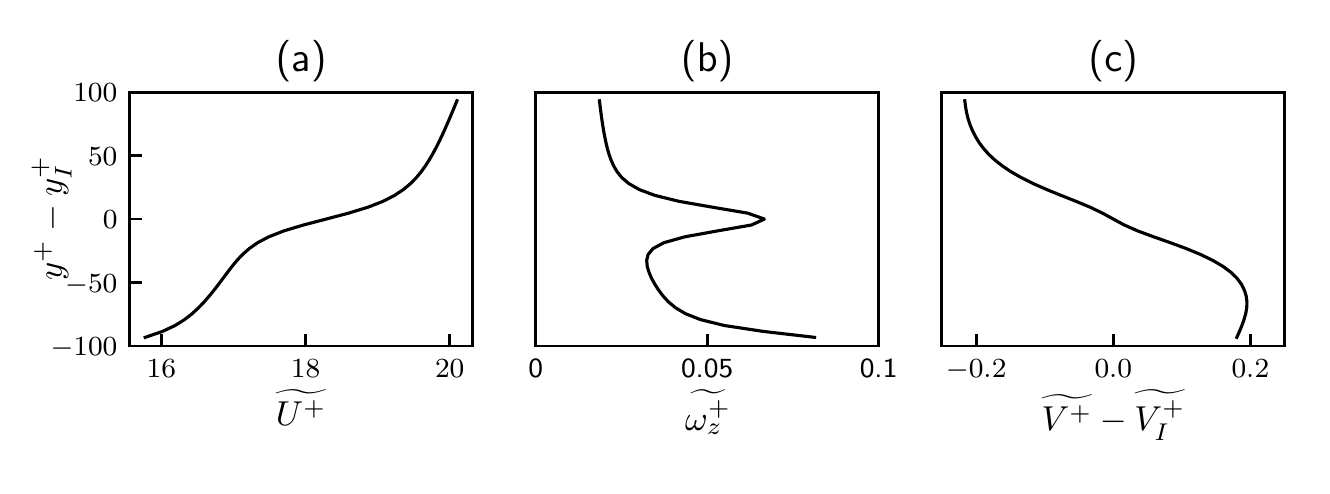}}
  \caption{Conditional average across the IIL between Zone 2 and 3 of the (a) streamwise velocity, (b) spanwise vorticity, and  (c) relative wall-normal velocity in the channel flow at \(Re_{\tau}=950\)}
  \label{fig:conditional}
\end{figure}

\begin{figure}
  \centerline{\includegraphics[width=\textwidth]{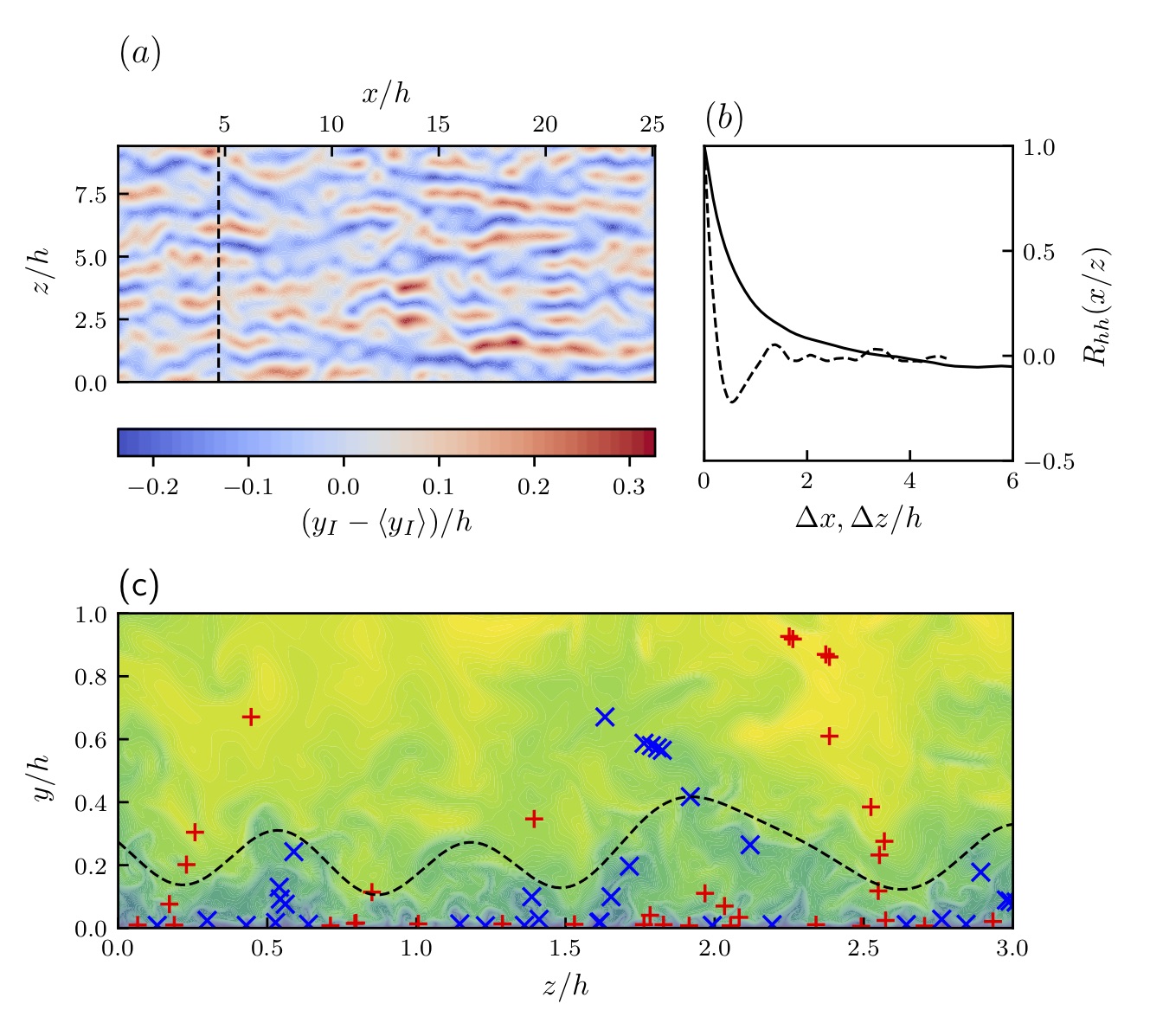}}
  \caption{(a) Fluctuations of the IIL (between zones 2 and 3)  normalized height. The raw data is  low-pass-filtered and re-sampled onto a coarser grid (\(\Delta x \times \Delta z =  0.1h\times 0.1h\)). The dashed vertical line represents the sliced plane investigated in figure (c);  (b) Streamwise solid and spanwise (dashed) autocorrelations of the IIL height fluctuation (computed using the raw data). (c) Comparison of the characteristic splines (CS) representing structures with a positive (\textbf{\textcolor{red}{+}}) and negative ($\textcolor{blue}{\times}$) streamwise velocity component in a single spanwise plane of the channel flow at \(Re_{\tau}=950\). The interfacial layer between zones 2 and 3 is overlaid. A low-pass filtering was done to remove the small scale oscillation of the IIL.}
  \label{fig:correlation}
\end{figure}

\modify{The probability distribution function of the IIL height, \(y_I\), for \(Re_{\tau}=550\) and \(950\) cases is shown in figure \ref{fig:distribution} (a)-(b). To a first order, the interfacial height has a log-normal  distribution with a mean equal to  \(\ln(Y_I/h)=\)-1.48/-1.53  and  standard deviation  of \(0.62\)/\(0.61\)  for the \(Re_{\tau}=550\)/\(950\) cases.  Further away from the wall, the height of the {TNTI} has  been reported to follow a Gaussian distribution in the turbulent boundary layer \citep{chauhan2014turbulent} and other flows \citep{bisset2002turbulent, hickey_hussain_wu_2013}. However, a positive skewness--\(0.645\)/0.641--is noted in the IIL height distribution. This implies, not too surprisingly, that the presence of the wall has greater influence in IIL than on the TNTI which has a skewness that is nearly zero. As a result, a log-normal fit as seen in figure \ref{fig:distribution}(a)-(b) provides a better description of the IIL height distribution.}

The comparison of the mean interfacial height in the \(Re_{\tau}=550\) and 950 cases suggests a $Re$-dependence. The normalized height of the interface decreases with \(Re_{\tau}\) a result which is consistent with \citet{kwon2014quiescent}'s observation that the thickness of quiescent core increase with \(Re_{\tau}\)\add{, this is further explored in section 3.4.}  The  mean location of the IIL normalized by viscous unit is overlaid on the mean streamwise velocity profile (see figure \ref{fig:distribution}(c)).  As noted earlier, zone 2 is associated with the logarithmic layer and the location of the IIL (between zones 2 and 3) defines  the top of logarithmic layer in the mean profile. The connection between IIL/UMZ and logarithmic layer has  been postulated by \citet{meinhart1995existence} but their detailed investigation was unable to provide evidence of this connection partially because of a lack of a robust and unambiguous IIL detection method. The present approach will allow us to address many of these lingering issues.

The conditional averages are computed across the interfacial layer (between zones 2 and 3) and  provide further statistical evidence of their importance. Here the conditional averages are computed at a relative distance from the interfacial layer, $y^+-y^+_I$ (where $y^+-y^+_I<0$ indicates a location below the interfacial layer). Figure \ref{fig:conditional}~(a) shows the large streamwise velocity jump across the interfacial layer which has been reported in previous works \citep{eisma2015interfaces,de2017interfaces}. At the IIL, a very clear peak in the spanwise vorticity is observed from the conditional statistics in figure \ref{fig:conditional}~(b).  Interestingly, the shear-like nature of the IIL may also result in a differential wall-normal velocity relative to the interface, as shown in figure \ref{fig:conditional}~(c). These conditional statistics support the understanding that the IIL acts as a strong shear layer in the wall turbulence.

Given the cluster algorithm's ability to unambiguously extract the three-dimensional interface,  the topology of the IIL is examined in the large spanwise domain, channel flow simulation. \add{We recall that the three-dimensional interface is not defined based on a streamwise iso-velocity contour (as is the case for the histogram approach), instead the fuzzy clustering allows a delineation that is more representative of the underlying physics.} A streamwise-aligned organization of the interfacial height is clearly visible in figure \ref{fig:correlation} (a). The autocorrelation function of the interfacial height in the streamwise (\(R_{hh}(x)\)) and spanwise direction (\(R_{hh}(z)\)) further support this visual observation (figure \ref{fig:correlation} (b)). The spanwise length scale is approximately \(0.5h\) whereas the streamwise length scale is more than \(3h\). The distribution and correlations of the height underscores a streak-like feature of the IIL \modify{which has been believed to} be related to the well-established streaky structure in the buffer layer. \citet{kwon2014quiescent} identified IIL as the \(0.95U_{CL}\) iso-surface (\(U_{CL}\) is the centre line  velocity) in the channel flow simulation at \(Re_{\tau}=1000\) and also observed similar streak-like features. These very long streamwise length scales support the observation of \citet{adrian2000vortex} that UMZs is the natural results of  LSMs or vortex packets.  

\add{The quantitative connection between the large and very-large scale motion (denoted as LSM and VLSM) of the near wall turbulence and the  three-dimensional IIL is investigated.   Qualitatively, the LSMs and VLSMs are easily identified through any classical near-wall turbulent visualization approaches. Here, we rely on the formal streak detection algorithm proposed by \citet{lee_lee_choi_sung_2014}, to quantitatively identify the near-wall LSMs and VLSMs in order to relate them to the three-dimensional IILs. The streak detection rests on sequential data filtering (Gaussian followed by a long-wavelength-pass filter) and structure detection algorithms which are detailed in the referenced paper \citep{lee_lee_choi_sung_2014}. We investigate the channel flow at \(Re_{\tau}=950\) by extracting the characteristic splines (CS) which are located within a characteristic plane (CP). We are able to identify the very-long positive and negative streamwise velocity regions in the flow. These splines are shown to be related to underlying characteristic structures, such as hairpin vortices \citep{lee_lee_choi_sung_2014}. Figure \ref{fig:correlation} (c) shows the location of the positive ($+$) and negative ($\times$) characteristic splines of a single spanwise cut, the cut location is identified in figure \ref{fig:correlation} (a). We overlay the corresponding interfacial height (between zones 2 and 3) identified through our clustering algorithm. Here we note a  clear correlation between the positive and negative streamwise structures and the interfacial layer. On average, when a positive streak is observed, the IIL is drawn in closer to the wall; when a negative streak is observed, the IIL is pushed away from the wall. This observation correlates with the ejection and sweep-type mechanisms that are typical for these type of near wall structures.  The ability to define three-dimensional interfaces supports the postulated relationship between the near wall structures and the interfacial layers in wall-bounded turbulence. }

\add{\subsection{High-Reynolds number DNS}}\label{sec:highRe}
\add{For the most part,  interfacial layers in high-Reynolds number flows have only been investigated using experimental data. The difficulty in obtaining near-wall experimental data with a sufficient spatial and temporal resolution over a large, three-dimensional field means that analysis of these IILs has been limited.  Furthermore, existing histogram-based approaches are ill-suited for use on non-homogeneous grid spacing, as the grid refinement artificially affects the probability distribution of the streamwise velocity. Here we study IILs in the turbulent channel flow by \citet{lee_moser_2015} which is computed at $Re_\tau\approx5200$ using a Fourier-Galerkin pseudo-spectral method in the streamwise and spanwise directions and a B-spline collocation method in the wall-normal direction. The details of the resolution are provided in table \ref{tab:piv} and further details can be found in \citet{lee_moser_2015}. Despite the high-Reynolds number of the simulation, only four UMZs are unambiguously identified using our KDE algorithm. A validation based on a manual approach using a classical histogram-based algorithm found a similar result.  This finding stands in contrast with some previous works suggesting a clear dependence of the Reynolds number  on the number  of uniform momentum zones \cite{de2017interfaces}--although previous results were conducted in boundary layer flows not channels. }

\add{The interfaces on a single slice of the data are shown by dashed lines in figure \ref{fig:zone_num_sens}. Here we note that the first identified layer is located very near the wall given the high Reynolds number of the flow. This observation is consistent with the location of the IILs as a function of the Reynolds number of the flow as suggested by  \citet{kwon2014quiescent}. As the FCM method is not rigidly delineating based on iso-velocity lines, we see that in some regions, UMZs may disappear.  For example, if we look at a vertical line from the wall at $x/h=2.8$, we traverse zones 1, 2, 3, and 4; a clear interface between the zones is observed. On the other hand, at around $x/h=3.0$, zone 2 is not present as the algorithm identifies only zones 1, 3, and 4 at that streamwise location; the interfaces between zones 1/2, and 2/3 are nearly colocated. This feature of the local adaption of the number of zones is consistent with the prevailing understanding of  IILs in turbulent wall bounded flows and presents a clear advantage for the study of high-Reynolds number flows.} \\

\par
\section{\add{Sensitivity and robustness of identification method}}\label{sec:robustness}
\add{The advantages of the Fuzzy Cluster Method (FCM) with respect to the standard histogram-based approach have been detailed in the previous sections and showed how the method can be extended to three-dimensional surface identification. Here we evaluate the sensitivity and robustness of the proposed IIL identification method. More specifically, we evaluate the sensitivity of the method to the \emph{a priori} selection of the number of zones, initialization of the clustering, and to a stochastic perturbation on the raw data (representing experimental error or numerical noise).}

\begin{figure}
    \centerline{\includegraphics[width=1.0\linewidth]{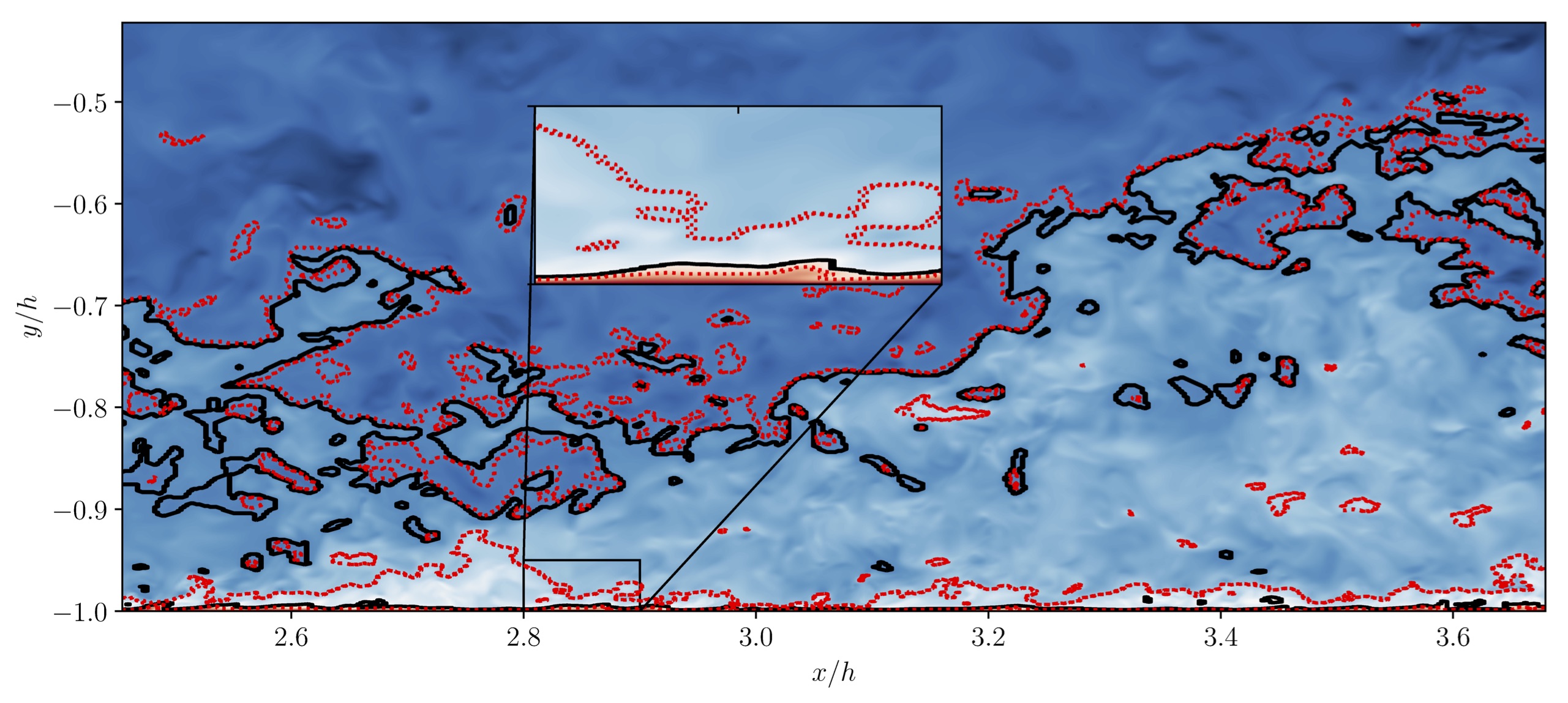}}
    \caption{The IIL overlaid on the streamwise velocity contour plot for a high Reynolds number channel flow \citep{lee_moser_2015}. The black solid lines are the identified IILs initialized with three internal zones and the red dotted lines are the identified IILs initialized with four internal zones. 
    \label{fig:zone_num_sens}}
\end{figure}

\add{\subsection{Sensitivity to the number of IILs}}
\add{The FCM requires a user-defined number of internal zones, $K$, as an \emph{a priori} input for clustering. In section \ref{sec:detection}, we proposed a method to automatically determine the number of internal zones in a given flow snapshot based on a Kernel Density Estimation (KDE). The question then arises, how sensitive is the FCM identification if the wrong number of internal zones, $K$, are provided as input?  We show that  the identification of the most important internal layers is insensitive to the number of zones if the error on $K$ is $\pm 1$.  Figure~\ref{fig:zone_num_sens} shows the IILs of a single slice of the high-Reynolds number channel flow DNS by \citet{lee_moser_2015} using three ($K=3$, full line) and four ($K=4$, dashed line) internal zones (with respectively two and three IILs). The same two IILs are identified using a different number of clusters: very near the wall and at the outer edge of the boundary layer). When clustering on four zones (three IILs) instead of three (two IILs), an additional interface is identified. In other words, the approach remains robust to a small error on the \emph{a priori} quantification of the number of zones. This same analysis was also conducted using fives UMZs with similar results. It should be noted that on lower Reynolds number datasets, the result showed a slightly greater sensitivity to the error. }

\begin{figure}
\centering
\begin{minipage}[c]{0.34\textwidth}
    \includegraphics[width=1.05\textwidth]{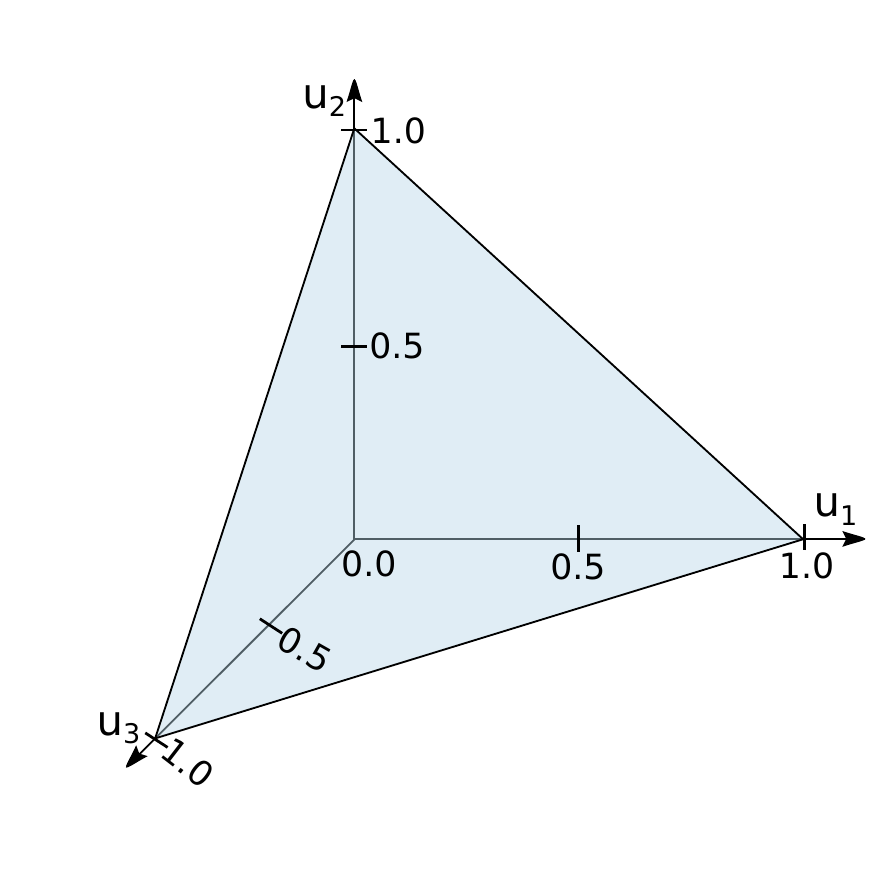}
  \end{minipage}
\begin{minipage}[c]{0.65\textwidth}
\includegraphics[width=\textwidth]{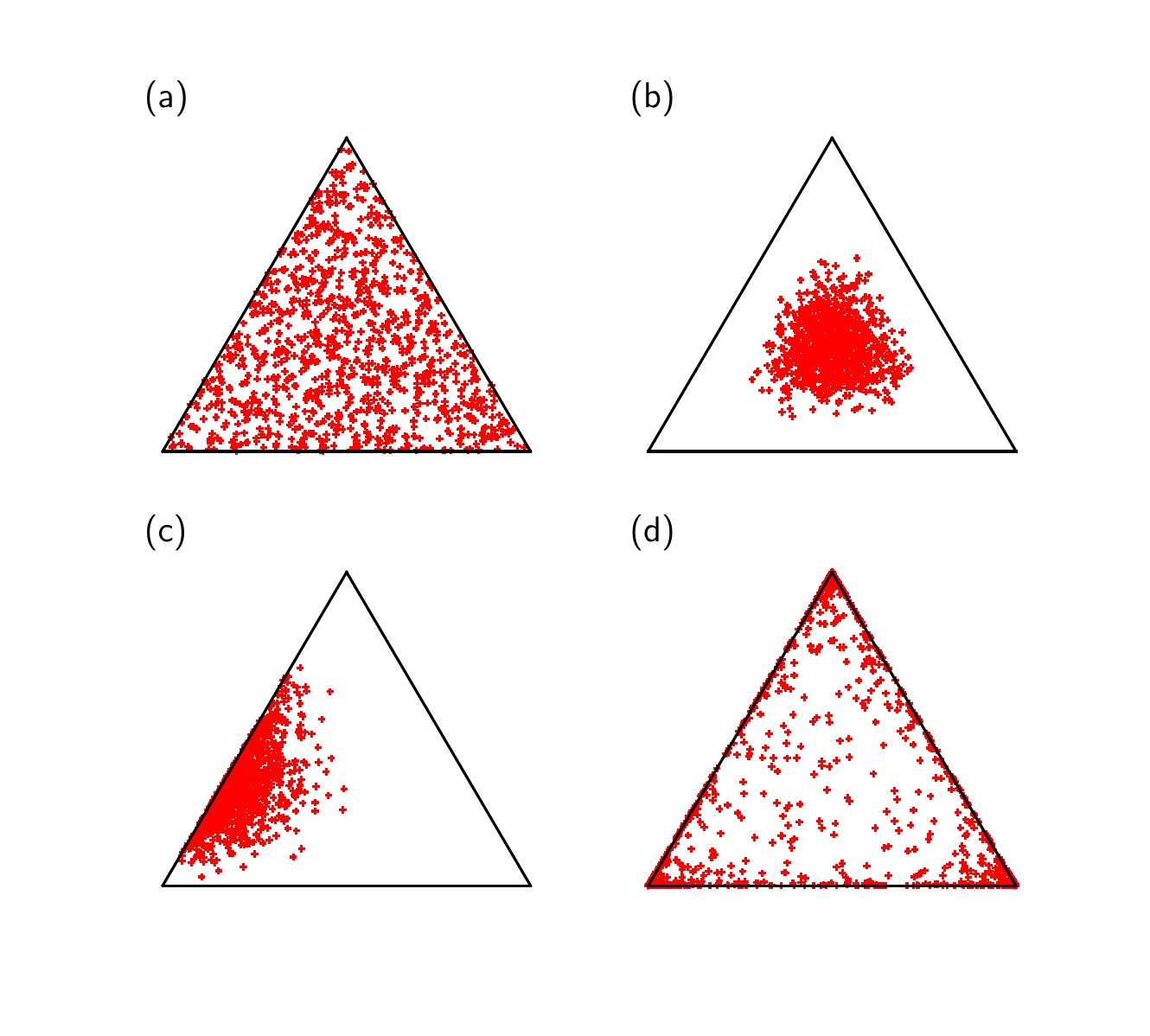}
\end{minipage}
    \caption{The Dirichlet distribution of the membership function initialization where each side of the triangle represents the initialization of one of the membership function. The points in the triangle must simultaneously satisfy two constraints: the value must be constrained between 0 and 1 in each dimension and must sum up to 1 (see illustration on left side).    The parameters used to defined the Dirichlet probability distribution  are $\alpha_1, \alpha_2, \alpha_3$. The above distributions represent (see Table 2):  (a)  Case 1; (b) Case 2; (c)     Case 3; and (d) Case 4.
    \label{fig:dirichlet}}
\end{figure}

\add{\subsection{Sensitivity to initialization}}
\add{One of the shortcomings of many data clustering approaches lies in the sensitivity of the final cluster to the initialization.  In order to initiate the fuzzy cluster algorithm, the membership functions, $u_{mk}$ must be initialized. The number of membership functions is the same as the \emph{a priori} determined number of clusters, $K$. Suppose we have three clusters ($K=3$), the sum of all of the membership functions at each $m$ grid points is:}
\begin{equation}
u_{m1}+u_{m2}+u_{m3}=1
\end{equation}
\add{To study the sensitivity of the FCM, we randomly select the initial membership function centroids in order to test the robustness of the method to the initialization. A three-dimensional Dirichlet probability distribution is used to randomly select the centroids for initialization. The Dirichlet distribution is fully defined by three independent parameters ($\alpha_1,\alpha_2,\alpha_3$), and under the proper normalization, the sum of the membership functions is unity. Figure \ref{fig:dirichlet} shows the tested Dirichlet probability distributions for initialization; each point in the triangle must have a value constrained between 0 and 1 in each dimension, and must sum up to one.  Irrespective of our selected initialization the exact same IILs (height, statistics and centroids) are identified for the channel flow simulation at $Re_\tau=950$ (centerplane slice). The statistics on the modal velocities, the normalized height of the interface ($\mu/h$) and its standard deviation $\sigma$/$h$ are shown in table 2. These results clearly show that the method remains robust to the initialization of the problem.}
\begin{table}
\begin{center}
\begin{tabular}{ccccccc}
& $\alpha_1$ & $\alpha_2$ & $\alpha_3$ & $\vec{\mathbf{c}}$ / $U_b$ & $\mu$/ $h$ & $\sigma$/$h$ \\
Case 1 & 1 & 1 & 1 & $(0.129,0.648,0.930)$ & 0.332515 & 0.108964 \\
Case 2 & 10 & 10 & 10 & $(0.129,0.648,0.930)$ & 0.332515 & 0.108964 \\
Case 3 & 1 & 5 & 10 & $(0.129,0.648,0.930)$ & 0.332515 & 0.108964 \\
Case 4 & 0.2 & 0.2 & 0.2 & $(0.129,0.648,0.930)$ & 0.332515 & 0.108964
\end{tabular}
\caption{Dirichlet distribution parameters, converged cluster centroid and statistics ($\mu$ and $\sigma$ is respectively the mean and standard deviation of the interfacial height) of IIL between Zone 2 and 3 for Case 1, 2, 3 and 4.}
\end{center}
\end{table}

\begin{figure}
    \centerline{\includegraphics[width=1.0\linewidth]{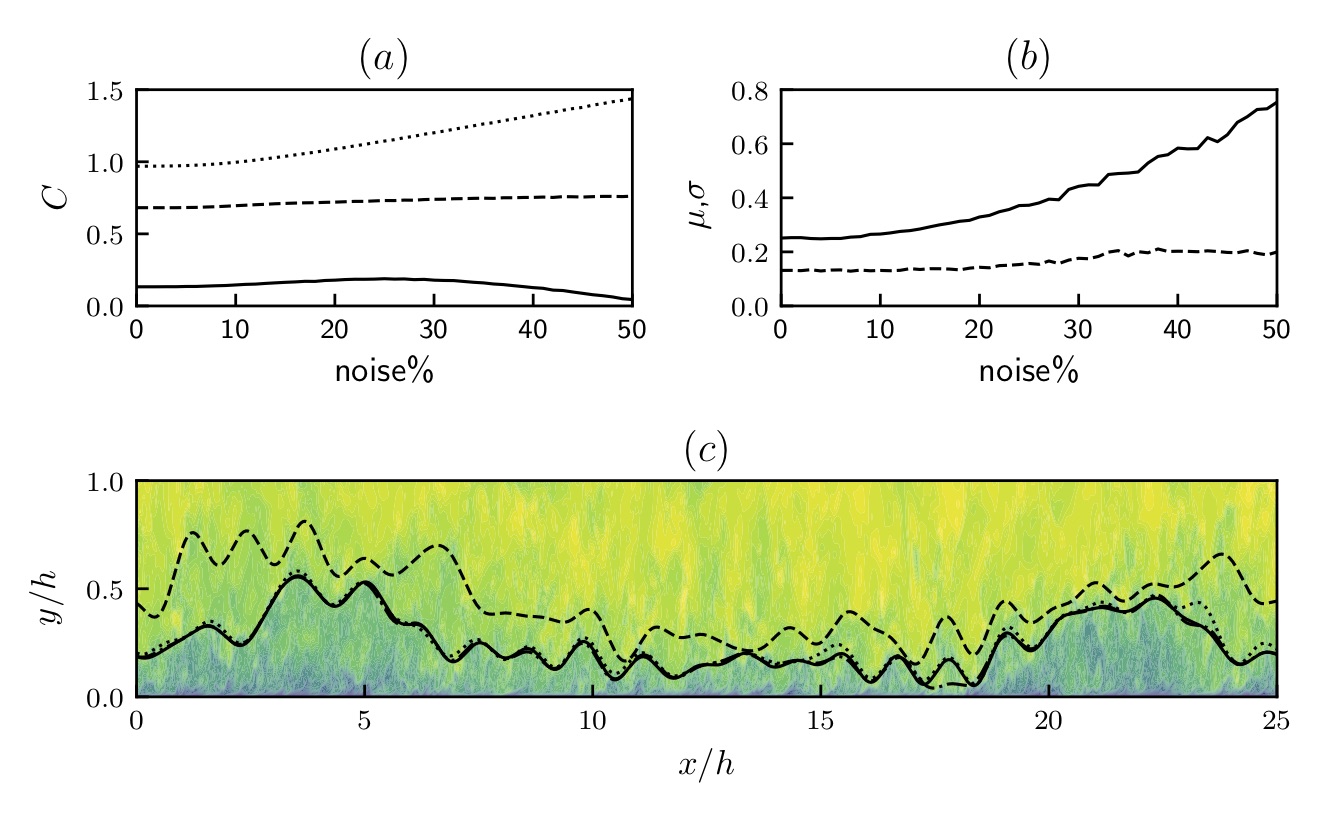}}
    \caption{\label{fig:perturbation}\add{Robustness of the FCM  against initial perturbations to the data, (a) shows the effect of white noise on the identified cluster centers; each line corresponds to the modal velocity of the UMZs. As the amplitude of the noise increases (above say 15\%), the ability to capture the cluster center decreases. (b) Shows the average height of the IIL between zones  2 and 3 (solid line) and the standard derivation (dashed line) as a function of the imposed noise to the data. (c) Shows the IIL between zones 2 and 3 under 0\% (solid), 5\% (dash-dotted), 10\% (dotted), and 30\% (dashed) white noise.}}
\end{figure}

\add{\subsection{IIL identification under stochastic perturbations}}
\add{As the clustering approach remains agnostic of the underlying physics of the flow, we test the robustness of the interfacial identification to initial perturbations to the data. Such perturbations could arise from experimental noise or discretization errors. Using the same initial membership functions (with a defined number of clusters), we perturb the raw data with white noise; here we use the channel flow at $Re_\tau=950$ for reporting the analysis, similar findings are observed on the other data.   We perturb the numerical data with up to 50\% white noise based on the bulk velocity of the flow. To show the robustness of the method, the perturbations are not damped down to the wall thus creating a non-physical noise. Figure \ref{fig:perturbation} (a) shows the effect of the white noise on the identification of the cluster centers (which correspond to the modal velocities of the UMZs); under 15\% noise, very little variation in the modal velocity is noted. The effect of the added noise on the average height of the IIL is presented in figure \ref{fig:perturbation} (b). The average height of the interface and its standard deviation remains constant up to 15\% white noise.  Figure \ref{fig:perturbation} (c) shows the actual interfacial identification between zone 2 and 3 under 0\% (solid), 5\% (dash-dotted), 10\% (dotted), and 30\% (dashed) noise. A very similar interface is captured  under all the lower intensity perturbations. At the unrealistically high level of 30\% noise, an error on the interface is observed. }\\

\section{Conclusion} \label{sec:conclusion}
We propose a novel cluster-based algorithm to identify interfacial layers in turbulent flows using a fuzzy cluster method (FCM). The approach is unambiguous, robust, and repeatable. Most importantly, it overcomes some of the limitations of the well-established histogram-based identification approaches. The new approach is invariant to the streamwise domain size, inhomogeneous near-wall grid spacing, and/or number of bins used in the histogram approach. Furthermore, it can be extended, without limitations, to three-dimensional flow fields. The only user-defined quantity is the number of uniform momentum zones (UMZs) to be extracted, which can be physically justified. \add{We also propose an algorithm based on a  Kernel Density Estimation (KDE) to automatically identify the number of UMZs for a given dataset; the KDE algorithm is only used for an \emph{a priori} estimation of the number of UMZs. We evaluated the sensitivity of the cluster algorithm to an error in the number of UMZs, initialization of the membership functions, and stochastic perturbations of raw data; the sensitivity results highlight the robustness of the proposed method.} The FCM was successfully applied to boundary layer  data (experimental, PIV)  and recovers the identified layers by the classical histogram- (IIL) and threshold-based (TNTI) approaches. The method identifies the interfaces for all datasets, whereas the histogram method only functions on about half the available PIV frames. As the proposed method can be extended to three-dimensions, the additional sampling data of the extracted IILs in a turbulent channel flow (numerical, DNS) is used to better understand the relationship between the interfacial layer and mean flow profile. We find that the mean interfacial height is located at the end of the log-layer and shows a strong Reynolds number dependence. The three-dimensional structures of the interface show a strong streamwise coherence, revealing a streaky-like structural pattern. \add{We show that the large- and very-large scale motion within the boundary is  correlated to the three-dimensional undulations of the IIL.} Finally, the conditional sampling reveals a very strong peak of spanwise vorticity at the interface. These proposed identification method will be applied to large-scale, transitional flow data by \citet{WuE5292} for further insight into the dynamics of interfacial layers.

\section*{Acknowledgments}
We acknowledge the support of the Natural Sciences and Engineering Research Council of Canada (NSERC) and the Natural Sciences Foundation of China (Grant No. 11472055). \add{The large-scale data post-processing was done with the support provided by SciNet (\url{www.scinethpc.ca}) and Compute Canada (\url{www.computecanada.ca})}.

\bibliographystyle{jfm}
\bibliography{Ref}

\begin{thebibliography}{30}
\expandafter\ifx\csname natexlab\endcsname\relax\def\natexlab#1{#1}\fi
\def\au#1{#1} \def\ed#1{#1} \def\yr#1{#1}\def\at#1{#1}\def\jt#1{\textit{#1}}
  \def\bt#1{#1}\def\bvol#1{\textbf{#1}} \def\vol#1{#1} \def\pg#1{#1}
  \def\publ#1{#1}\def\arxiv#1{#1}\def\org#1{#1}\def\st#1{\textit{#1}}

\bibitem[Adrian {\em et~al.\/}(2000)Adrian, Meinhart \&
  Tomkins]{adrian2000vortex}
{\sc \au{Adrian, R.~J.}, \au{Meinhart, C.~D.} \& \au{Tomkins, C.~D.}} \yr{2000}
   \at{Vortex organization in the outer region of the turbulent boundary
  layer}.  \jt{Journal of Fluid Mechanics}  \bvol{422},  \pg{1--54}.

\bibitem[Bezdek(1974)]{bezdek1974numerical}
{\sc \au{Bezdek, J.~C.}} \yr{1974}  \at{Numerical taxonomy with fuzzy sets}.
  \jt{Journal of Mathematical Biology}  \bvol{1}~(1),  \pg{57--71}.

\bibitem[Bezdek(1981)]{bezdek1981objective}
{\sc \au{Bezdek, J.~C.}} \yr{1981}  \at{Objective function clustering}.  \bt{In
  {\em Pattern recognition with fuzzy objective function algorithms\/}},
  \pg{pp. 43--93}.  \publ{Springer}.

\bibitem[Bisset {\em et~al.\/}(2002)Bisset, Hunt \&
  Rogers]{bisset2002turbulent}
{\sc \au{Bisset, D.~K.}, \au{Hunt, J. C.~R.} \& \au{Rogers, M.~M.}} \yr{2002}
  \at{The turbulent/non-turbulent interface bounding a far wake}.  \jt{Journal
  of Fluid Mechanics}  \bvol{451},  \pg{383--410}.

\bibitem[Borrell \& Jim{\'e}nez(2016)]{borrell2016properties}
{\sc \au{Borrell, G.} \& \au{Jim{\'e}nez, J.}} \yr{2016}  \at{Properties of the
  turbulent/non-turbulent interface in boundary layers}.  \jt{Journal of Fluid
  Mechanics}  \bvol{801},  \pg{554--596}.

\bibitem[Chauhan {\em et~al.\/}(2014)Chauhan, Philip, de~Silva, Hutchins \&
  Marusic]{chauhan2014turbulent}
{\sc \au{Chauhan, K.}, \au{Philip, J.}, \au{de~Silva, C.~M.}, \au{Hutchins, N.}
  \& \au{Marusic, I.}} \yr{2014}  \at{The turbulent/non-turbulent interface and
  entrainment in a boundary layer}.  \jt{Journal of Fluid Mechanics}
  \bvol{742},  \pg{119--151}.

\bibitem[Corrsin \& Kistler(1955)]{corrsin1955free}
{\sc \au{Corrsin, S.} \& \au{Kistler, A.~L.}} \yr{1955}  \bt{{Free-stream
  boundaries of turbulent flows}}. {\em Tech. Rep.\/}.  \org{Johns Hopkins
  University (NACA-TR-1244)}.

\bibitem[Del~{\'A}lamo {\em et~al.\/}(2004)Del~{\'A}lamo, Jim{\'e}nez,
  Zandonade \& Moser]{del2004scaling}
{\sc \au{Del~{\'A}lamo, J.~C.}, \au{Jim{\'e}nez, J.}, \au{Zandonade, P.} \&
  \au{Moser, R.~D.}} \yr{2004}  \at{Scaling of the energy spectra of turbulent
  channels}.  \jt{Journal of Fluid Mechanics}  \bvol{500},  \pg{135--144}.

\bibitem[Dunn(1973)]{dunn1973fuzzy}
{\sc \au{Dunn, J.~C.}} \yr{1973}  \at{A fuzzy relative of the isodata process
  and its use in detecting compact well-separated clusters}.  \jt{Cybernetics
  and Systems}  \bvol{3}~(3),  \pg{32--57}.

\bibitem[Eisma {\em et~al.\/}(2015)Eisma, Westerweel, Ooms \&
  Elsinga]{eisma2015interfaces}
{\sc \au{Eisma, J.}, \au{Westerweel, J.}, \au{Ooms, G.} \& \au{Elsinga, G.~E.}}
  \yr{2015}  \at{Interfaces and internal layers in a turbulent boundary layer}.
   \jt{Physics of Fluids}  \bvol{27}~(5),  \pg{055103}.

\bibitem[Friedman {\em et~al.\/}(2001)Friedman, Hastie \&
  Tibshirani]{friedman2001elements}
{\sc \au{Friedman, J.}, \au{Hastie, T.} \& \au{Tibshirani, R.}} \yr{2001} {\em
  The elements of statistical learning\/}.  \publ{Springer}.

\bibitem[Hickey {\em et~al.\/}(2013)Hickey, Hussain \&
  Wu]{hickey_hussain_wu_2013}
{\sc \au{Hickey, J.}, \au{Hussain, F.} \& \au{Wu, X.}} \yr{2013}  \at{Role of
  coherent structures in multiple self-similar states of turbulent planar
  wakes}.  \jt{Journal of Fluid Mechanics}  \bvol{731},  \pg{312–363}.

\bibitem[Holzner \& L{\"u}thi(2011)]{holzner2011laminar}
{\sc \au{Holzner, M.} \& \au{L{\"u}thi, B.}} \yr{2011}  \at{Laminar superlayer
  at the turbulence boundary}.  \jt{Physical review letters}  \bvol{106}~(13),
  \pg{134503}.

\bibitem[Kolar(2007)]{kolar2007vortex}
{\sc \au{Kolar, V.}} \yr{2007}  \at{Vortex identification : New requirements
  and limitations}.  \jt{International Journal of Heat and Fluid Flow}
  \bvol{28}~(4),  \pg{638--652}.

\bibitem[Kwon {\em et~al.\/}(2014)Kwon, Philip, de~Silva, Hutchins \&
  Monty]{kwon2014quiescent}
{\sc \au{Kwon, Y.~S.}, \au{Philip, J.}, \au{de~Silva, C.~M.}, \au{Hutchins, N.}
  \& \au{Monty, J.~P.}} \yr{2014}  \at{The quiescent core of turbulent channel
  flow}.  \jt{Journal of Fluid Mechanics}  \bvol{751},  \pg{228--254}.

\bibitem[Lee {\em et~al.\/}(2014)Lee, Lee, Choi \&
  Sung]{lee_lee_choi_sung_2014}
{\sc \au{Lee, Jin}, \au{Lee, Jae~Hwa}, \au{Choi, Jung-Il} \& \au{Sung,
  Hyung~Jin}} \yr{2014}  \at{Spatial organization of large- and
  very-large-scale motions in a turbulent channel flow}.  \jt{Journal of Fluid
  Mechanics}  \bvol{749},  \pg{818–840}.

\bibitem[Lee \& Moser(2015)]{lee_moser_2015}
{\sc \au{Lee, Myoungkyu} \& \au{Moser, Robert~D.}} \yr{2015}  \at{Direct
  numerical simulation of turbulent channel flow up to
  $\mathit{Re}_{{\it\tau}}\approx 5200$}.  \jt{Journal of Fluid Mechanics}
  \bvol{774},  \pg{395–415}.

\bibitem[Marusic {\em et~al.\/}(2013)Marusic, Monty, Hultmark \&
  Smits]{marusic_monty_hultmark_smits_2013}
{\sc \au{Marusic, I.}, \au{Monty, J.~P.}, \au{Hultmark, M.} \& \au{Smits,
  A.~J.}} \yr{2013}  \at{On the logarithmic region in wall turbulence}.
  \jt{Journal of Fluid Mechanics}  \bvol{716},  \pg{R3}.

\bibitem[Meinhart \& Adrian(1995)]{meinhart1995existence}
{\sc \au{Meinhart, C.~D.} \& \au{Adrian, R.~J.}} \yr{1995}  \at{On the
  existence of uniform momentum zones in a turbulent boundary layer}.
  \jt{Physics of Fluids}  \bvol{7}~(4),  \pg{694--696}.

\bibitem[Pal \& Bezdek(1995)]{Pal:1995:CVF:2234581.2235722}
{\sc \au{Pal, N.~R.} \& \au{Bezdek, J.~C.}} \yr{1995}  \at{On cluster validity
  for the fuzzy c-means model}.  \jt{Trans. Fuz Sys.}  \bvol{3}~(3),
  \pg{370--379}.

\bibitem[Pham {\em et~al.\/}(2000)Pham, Xu \& Prince]{pham2000current}
{\sc \au{Pham, D.~L.}, \au{Xu, C.} \& \au{Prince, J.~L.}} \yr{2000}
  \at{Current methods in medical image segmentation}.  \jt{Annual review of
  biomedical engineering}  \bvol{2}~(1),  \pg{315--337}.

\bibitem[Saxton-Fox \& McKeon(2017)]{saxton2017coherent}
{\sc \au{Saxton-Fox, T.} \& \au{McKeon, B.~J.}} \yr{2017}  \at{Coherent
  structures, uniform momentum zones and the streamwise energy spectrum in
  wall-bounded turbulent flows}.  \jt{Journal of Fluid Mechanics}  \bvol{826}.

\bibitem[Scott(1992)]{scott1992multivariate}
{\sc \au{Scott, David~W.}} \yr{1992} {\em Multivariate density estimation:
  theory, practice, and visualization\/}.  \publ{John Wiley \& Sons}.

\bibitem[de~Silva {\em et~al.\/}(2017)de~Silva, Philip, Hutchins \&
  Marusic]{de2017interfaces}
{\sc \au{de~Silva, C.M.}, \au{Philip, J.}, \au{Hutchins, N.} \& \au{Marusic,
  I.}} \yr{2017}  \at{Interfaces of uniform momentum zones in turbulent
  boundary layers}.  \jt{Journal of Fluid Mechanics}  \bvol{820},
  \pg{451--478}.

\bibitem[da~Silva {\em et~al.\/}(2014)da~Silva, Hunt, Eames \&
  Westerweel]{Silva2014}
{\sc \au{da~Silva, C.~B.}, \au{Hunt, J.~C.R.}, \au{Eames, I.} \&
  \au{Westerweel, J.}} \yr{2014}  \at{Interfacial layers between regions of
  different turbulence intensity}.  \jt{Annual Review of Fluid Mechanics}
  \bvol{46}~(1),  \pg{567--590}.

\bibitem[de~Silva {\em et~al.\/}(2016)de~Silva, Hutchins \&
  Marusic]{de2016uniform}
{\sc \au{de~Silva, C.~M.}, \au{Hutchins, N.} \& \au{Marusic, I.}} \yr{2016}
  \at{Uniform momentum zones in turbulent boundary layers}.  \jt{Journal of
  Fluid Mechanics}  \bvol{786},  \pg{309--331}.

\bibitem[Silverman(1998)]{silverman1998density}
{\sc \au{Silverman, Bernard~W.}} \yr{1998} {\em Density estimation for
  statistics and data analysis\/}.  \publ{Routledge}.

\bibitem[Tomkins {\em et~al.\/}(1998)Tomkins, Adrian \&
  Balachandar]{tomkins1998structure}
{\sc \au{Tomkins, C.}, \au{Adrian, R.} \& \au{Balachandar, S.}} \yr{1998} The
  structure of vortex packets in wall turbulence.  \bt{In {\em 29th AIAA, Fluid
  Dynamics Conference\/}},  \pg{p. 2962}.

\bibitem[Westerweel {\em et~al.\/}(2005)Westerweel, Fukushima, Pedersen \&
  Hunt]{westerweel2005mechanics}
{\sc \au{Westerweel, J}, \au{Fukushima, C}, \au{Pedersen, JM} \& \au{Hunt,
  JCR}} \yr{2005}  \at{Mechanics of the turbulent-nonturbulent interface of a
  jet}.  \jt{Physical review letters}  \bvol{95}~(17),  \pg{174501}.

\bibitem[Wu {\em et~al.\/}(2017)Wu, Moin, Wallace, Skarda, Lozano-Dur{\'a}n \&
  Hickey]{WuE5292}
{\sc \au{Wu, X.}, \au{Moin, P.}, \au{Wallace, J.~M.}, \au{Skarda, J.},
  \au{Lozano-Dur{\'a}n, A.} \& \au{Hickey, J.-P.}} \yr{2017}
  \at{Transitional{\textendash}turbulent spots and
  turbulent{\textendash}turbulent spots in boundary layers}.  \jt{Proceedings
  of the National Academy of Sciences}  \bvol{114}~(27),  \pg{E5292--E5299}.

\end{thebibliography}
\end{document}